\documentclass[12pt]{article}
\usepackage{hyperref}
\usepackage{amsmath}
\usepackage{graphicx}
\usepackage{amssymb} 
\usepackage{color}
\definecolor{mypurple}{rgb}{0.71,0.02,1}

\def\be{\begin{equation}}
\def\ee{\end{equation}}
\def\bea{\begin{eqnarray}}
\def\eea{\end{eqnarray}}
\def\bi{\begin{itemize}}
\def\ei{\end{itemize}}
\def\dd{\mathrm{d}}

\date{}

\title{ Four-vector vs. four-scalar representation of the Dirac wave function}

\author{
Mayeul Arminjon\,$^{1}$ and Frank Reifler\,$^2$\\
$^1$ \small\it Laboratory ``Soils, Solids, Structures, Risks'' (CNRS, UJF, and G-INP),\\
\small\it BP 53, F-38041 Grenoble cedex 9, France.\\
\small\it $^2$ Lockheed Martin Corporation, MS2 137-205,\\ 
\small\it 199 Borton Landing Road, Moorestown, New Jersey 08057, USA.
} 

\begin{document}
\maketitle

\begin{abstract}
\noindent In a Minkowski spacetime, one may transform the Dirac wave function under the spin group, as one transforms coordinates under the Poincar\'e group. This is not an option in a curved spacetime. Therefore, in the equation proposed independently by Fock and Weyl, the four complex components of the Dirac wave function transform as scalars under a general coordinate transformation. Recent work has shown that a covariant complex four-vector representation is also possible. Using notions of vector bundle theory, we describe these two representations in a unified framework. We prove theorems that relate together the different representations and the different choices of connections within each representation. As a result, either of the two representations can account for a variety of inequivalent, linear, covariant Dirac equations in a curved spacetime that reduce to the original Dirac equation in a Minkowski spacetime. In particular, we show that the standard Dirac equation in a curved spacetime, with any choice of the tetrad field, is equivalent to a particular realization of the covariant Dirac equation for a complex four-vector wave function.\\

\vspace{30mm}
\end{abstract}

\section{Introduction}\label{Introduction}

The original Dirac equation applies in the Minkowski spacetime of special relativity. As the coordinates are Lorentz-transformed, the Dirac wave function transforms under the spin group. However, Weyl \cite{Weyl1929b} and Fock \cite{Fock1929b} recognized that, on changing the coordinates, transforming the Dirac wave function under the spin group is not an option in a curved spacetime, or even in a flat spacetime with affine coordinates. They proposed independently what has become the standard version of the Dirac equation in a curved spacetime \cite{Weyl1929b,Fock1929b}, hereafter the Dirac-Fock-Weyl ({\it DFW}) equation. As is well known for the DFW equation  \cite{BrillWheeler1957+Corr, deOliveiraTiomno1962, ChapmanLeiter1976}, the four complex components of the wave function transform as a scalars under a general coordinate transformation. Recently, two alternative extensions of the Dirac equation to a curved spacetime have been proposed \cite{A39}, based on the {\it tensor representation of the Dirac field} ({\it TRD}) \cite{A39,A37}. In these alternative equations, the wave function is a complex four-vector and the set of the components of the four Dirac matrices builds a third-order affine tensor. Thus, there are only two possibilities in a curved spacetime. Either the Dirac wave function transforms as a quadruplet of four complex scalar fields under coordinate transformations as in DFW, or it transforms as a complex four-vector field as in TRD. In a flat spacetime with linear affine coordinates, and constant Dirac matrices, TRD and DFW have been shown to be equivalent \cite{A40}. Thus there can be no question as to the fermion content of TRD.\\

On the other hand, locally, in a curved spacetime or in a flat spacetime with general coordinates, the set of four complex $4\times 4$ Dirac matrices $\gamma^\mu(X)$ depends on the point $X$ in spacetime, thus it becomes a field $X \mapsto \gamma^\mu(X)$, for which there is a continuum of different possible choices---all satisfying the same anticommutation relation in the given Lorentzian spacetime $(\mathrm{V},g_{\mu\nu})$:
\be \label{Clifford}
\gamma ^\mu \gamma ^\nu + \gamma ^\nu \gamma ^\mu = 2g^{\mu \nu}\,{\bf 1}_4, \quad \mu ,\nu \in \{0,...,3\} \quad ({\bf 1}_4\equiv \mathrm{diag}(1,1,1,1)).
\ee
At any point $X$ in the spacetime, any two possible choices $\gamma^\mu(X)$ and $\widetilde{\gamma}^\mu(X)$ are related together by a local similarity transformation $S(X) \in {\sf GL(4, C)}$, which is unique up to a non-zero complex factor $\lambda(X)$, such that we have \cite{A40,Pauli1936}:
\be \label{similarity-gamma}
\widetilde{\gamma} ^\mu =  S^{-1}\gamma ^\mu S, \quad \mu =0,...,3.
\ee
For the DFW equation, the Dirac matrices $\gamma^\mu$ are defined through an orthonormal tetrad field \cite{BrillWheeler1957+Corr, deOliveiraTiomno1962, ChapmanLeiter1976}. This implies, as is well known, that only the local similarity transformations which at any point belong to the spin group are admissible. It is also well known that the DFW equation is covariant \cite{BrillWheeler1957+Corr, ChapmanLeiter1976} under any such (differentiable) admissible local similarity, $X \mapsto S(X) \in {\sf Spin(1,3)}$, when this is simultaneously applied to the gamma field by Eq. (\ref{similarity-gamma}), and to the wave function by
\be\label{psitilde=S^-1 psi} 
\widetilde{\Psi}=S^{-1}\Psi.
\ee
For TRD, the $\gamma^\mu$ field is not necessarily defined through an orthonormal tetrad field. Accordingly, the similarity matrices $S(X)$ can then be any element of the linear group ${\sf GL(4, C)}$ \cite{A42}. In the literature, it is assumed that the covariance of the DFW equation under the admissible similarities implies a complete physical insensitivity to the different possible choices of the  $\gamma^\mu$ field. (This assumption was stated explicitly by Audretsch \cite{Audretsch1974p328}.) However, it turns out that the Hamiltonian operator H in a given coordinate system does depend on the $\gamma^\mu$ field for both TRD and DFW theories \cite{A42,A43,B31}. For instance, the hermiticity of the Dirac Hamiltonian H is not preserved under all local similarity transformations that are admissible for DFW, implying that the validity of Leclerc's hermiticity condition for H \cite{Leclerc2006} is not general \cite{A42}. (For more about the Hamiltonian theory see Ref. \cite{A42}.) \\

Furthermore, unlike in DFW, the $\gamma^\mu$ 's in TRD are not required to be covariantly constant. Since the choice of the Dirac matrices $\gamma^\mu$ is less constrained in TRD theory than in DFW theory, it might be the case that each of the TRD equations, say TRD--1 and TRD--2, be {\it more general} than the standard DFW equation. The main aim of this paper is to prove that this is indeed the case in a precise sense. We will prove that, in any given non-compact, four-dimensional, Lorentzian spacetime $(\mathrm{V},g_{\mu\nu})$ admitting a spinor structure, any DFW equation, obtained by a particular choice of the tetrad field in that spacetime, is equivalent to a particular case of the TRD--1 equation (and also to a particular case of the TRD--2 equation). We will also prove that any DFW equation is equivalent to a particular case (amounting to a specific choice of the Dirac matrices) of a very simple form (``QRD--0'') of the Dirac equation in a curved spacetime, which is obtained by setting the connection matrices equal to {\it zero.} \\

We shall begin in Section \ref{Two representations} with a unified discussion of the two possible representations (QRD and TRD) in a non-compact, four-dimensional, Lorentzian spacetime admitting a spinor structure. This discussion is based on the introduction of spin-half fields, defined as sections of vector bundles. We shall also provide the link with matrix notation which is more commonly used in the physics literature. Then in Section \ref{SpecialClasses}, we shall introduce different classes of Dirac equations, including the simple form QRD--0, and a teleparallel version (also new). Section \ref{LocalSimilarities} will discuss the notion of a local similarity, which can be regarded as either a passive change of basis of the fibers of the vector bundles, or an active transformation of the spin-half fields. The Lagrangian common to all versions will be introduced in Section \ref{modified Dirac equation}. Then in Section \ref{Relations}, we will state and prove the theorems that establish equivalences between classes of Dirac equations.

\section{The two possible representations}\label{Two representations}

\subsection{A common geometrical framework}

Let U be an open subset of the spacetime V where local coordinates are defined and let $X \mapsto \gamma^\mu(X)\ \ (\mu=0,...,3)$ be a set of Dirac matrix fields
\footnote{\
In this subsection, we will use the terms ``matrix" and ``linear map" synonymously, noting that a matrix acting on a fiber of a vector bundle E is a linear map described in terms of some basis.  
}
defined on the complex tangent bundle $\mathrm{T}_{\sf C}\mathrm{U}$, satisfying the anticommutation relation (\ref{Clifford}).  Here ${\sf C}$ denotes the set of complex numbers; whereas, ${\sf C}^4$ will be the standard complex vector space consisting of quadruplets of complex numbers.  Furthermore, TU denotes the tangent bundle of U and $\mathrm{T}_{\sf C}\mathrm{U}$ denotes the complex tangent bundle of U. \\  

\noindent Then, corresponding to each vector $p \in \mathrm{TU}$ there is a Dirac matrix of the form: ${\not}p \equiv p^\mu \gamma_\mu$.  Note that the dagger notation is due to Feynman \cite{FeynmanQED}.  That is, there is a vector bundle map $\mathrm{TU} \rightarrow Hom\left(\mathrm{T}_{\sf C}\mathrm{U},\mathrm{T}_{\sf C}\mathrm{U}\right)$, taking $p \in \mathrm{TU}$ to ${\not}p \in Hom\left(\mathrm{T}_{\sf C}\mathrm{U},\mathrm{T}_{\sf C}\mathrm{U}\right)$.  Then, for all $p,k \in \mathrm{TU}$, it follows from the anticommutation relation (\ref{Clifford}) that \cite{FeynmanQED}:  
\be\label{Clifford slash-T_C U}
{\not}p{\not}k+{\not}k{\not}p=2<p,k>\mathrm{Id}_{\mathrm{T}_{\sf C}\mathrm{U}},
\ee
where $<\,,\,>$ is the spacetime metric and $\mathrm{Id}_{\mathrm{T}_{\sf C}\mathrm{U}}$ is the identity element of $Hom(\mathrm{T}_{\sf C}\mathrm{U},\mathrm{T}_{\sf C}\mathrm{U})$. Note that both Feynman's dagger notation ${\not}p$ as well as the anticommutation relation (\ref{Clifford slash}) are coordinate free.\\   

We will say that a complex vector bundle E on the spacetime V is a ``spinor bundle" if there is a global field of such Dirac matrices acting irreducibly on it.
\footnote{\
In this paper, all vector bundles will be smoothly defined over the spacetime V (or an open subset thereof which will be clear from the context) as their common base space. Sections of vector bundles will be smooth as well. A map between vector bundles, or {\it vector bundle map,} will be a (smooth) {\it morphism of vector bundles} which projects to the identity map on their common base space V (or open subset thereof) (\cite{Husemoller}, pp. 65--67; \cite{DieudonnéTome3}, paragraph 16.15.2).\\
}
More precisely we define:\\

\paragraph{Definition.}\label{DefSpinorBundle} {\it A smooth complex vector bundle $\mathrm{E}$, whose fiber is ${\sf C}^4$ and whose base space is a four-dimensional spacetime $\mathrm{V}$, will be called a spinor bundle if and only if there is a smooth vector bundle map: $\mathrm{TV} \rightarrow Hom\left(\mathrm{E},\mathrm{E}\right)$, which projects to the identity map on $\mathrm{V}$, taking each vector $p \in \mathrm{TV}$ to  $ {\not}p \in Hom\left(\mathrm{E},\mathrm{E}\right)$, satisfying the anticommutation relation} 
\be\label{Clifford slash}
{\not}p{\not}k+{\not}k{\not}p = 2<p,k>\,\mathrm{Id_E}.
\ee  
We note that this definition is equivalent to that given by Trautman \cite{Trautman2008} (Definition 2 on p. 247).
\footnote{\
To quote Trautman (Ref. \cite{Trautman2008}, p. 247): ``From the universality of Clifford algebras it follows that, to define $\tau $ [the morphism considered by Trautman], it is enough to give the restriction of $\tau $ to $T M \subset  Cl(g)$, this restriction being subject to $\tau  (v)^2 = g(v, v) id_{\Sigma _x}$ for every $v \in T_x M$."
}\\

Then, given any connection $D$ defined on the complex vector bundle E, the Dirac equation for a fermion particle of mass $m$   (setting Planck's constant $\hbar=1$ and the speed of light $c=1$) can be written for a smooth section $\psi$ of E  as follows:
\be\label{Dirac slash}
{\not}\mathcal{D}\psi = -im\psi,
\ee
where the global Dirac operator ${\not}\mathcal{D}$ is defined as \cite{Booss-Bavnbek}:
\be\label{Dirac operator slash}
{\not} \mathcal{D} \equiv \sum_{\alpha,\beta} \eta^{\alpha\beta} {\not} u_\beta D_{u_\alpha}
\ee
where $\eta^{\alpha\beta}\ (\alpha,\beta=0,...,3 )$ is the Minkowski metric, and $(u_\alpha)$ is any orthonormal basis of the tangent space $\mathrm{TV}_X$, at each spacetime point  $X$.  Note that Eq. (\ref{Dirac operator slash}) does not depend on the choice of orthonormal basis $(u_\alpha)$ chosen arbitrarily at each spacetime point $X$. \\

In order to define a Lagrangian, it is commonly assumed in the definition of
``spinor bundle" the further property that there exists a nonsingular, not
necessarily positive, Hermitian metric, smoothly defined on each fiber of the
complex vector bundle E, with respect to which the Dirac matrices are
Hermitian \cite{A40,Booss-Bavnbek}. This Hermitian metric was first introduced by Pauli \cite{Pauli1936,Pauli1933} and called a ``hermitizing" metric or matrix for the Dirac matrices. Denoting the hermitizing metric by $(\, ,\, )$, the Dirac Lagrangian is defined globally on the spacetime V as follows:
\be\label{Lagrangian-intrinsic}
L=
\ \frac{i}{2}
\left [(\psi,{\not}\mathcal{D}\psi)-({\not}\mathcal{D}\psi,\psi)+2im(\psi ,\psi) \right].
\ee
This Lagrangian depends on the choice of Dirac matrices and hermitizing
metric, as well as the choice of connection on the complex vector bundle E.
Note that similar to the spacetime metric $<\,,\,>$ which induces a canonical real linear
isomorphism from the tangent bundle TV to its dual TV$^\circ$, the hermitizing
metric $(\, ,\, )$ induces a canonical conjugate linear isomorphism from E to its
dual E$^\circ$. This conjugate linear isomorphism $\mathrm{E}\rightarrow \mathrm{E}^\circ$ which takes $\psi \rightarrow \overline{\psi }$ is called the ``spinor adjoint map". Further study of the Lagrangian (\ref{Lagrangian-intrinsic}) and the linear, covariant Dirac equations derived from it, for general choices of the Dirac matrices and hermitizing metrics and connections on E, is presented in Section \ref{modified Dirac equation}.\\

Recall that a spinor structure is defined to be a bundle map from a principal bundle $\mathcal{F}$, having a structure group ${\sf Spin(1,3)}$, to the bundle $\mathcal{O} $ of the orthonormal bases of the tangent bundle TV, such that some diagram commutes. (See e.g. Refs. \cite{Trautman2008, Geroch1968, Isham1978}. The bundle $\mathcal{O} $ is a principal bundle with structure group the special Lorentz group ${\sf SO(1,3)}$.) If the spacetime V is four-dimensional, noncompact, and admits a spinor structure in that sense, then there exists a global tetrad field on V \cite{Geroch1968}. The spacetime V is then said to be parallelizable. 
\footnote{\ 
In this paper, tetrad fields refer to smooth orthonormal frame fields of the tangent
bundle. Frame fields of an arbitrary vector bundle are assumed to be non-singular, but
not necessarily orthonormal, unless specified as such. Vector bundles, for which at least one smooth global frame field exists, are said to be parallelizable. A spacetime is said to be parallelizable if its tangent bundle is parallelizable.
}
According to Penrose and Rindler \cite{PenroseRindler1986}, these are the only spacetimes of physical interest. Such a spacetime V enjoys the following two properties: i) the trivial vector bundle $\mathrm{V} \times {\sf C}^4$ is a spinor bundle, and ii) the complex tangent bundle $\mathrm{T}_{\sf C}\mathrm{V}$ is a spinor bundle.  (See Appendix \ref{Construc-gamma} for a straightforward constructive proof.)\\ 

Properties (i) and (ii) motivate defining two representations of wave functions $\psi$.  In the first representation of a wave function, used in the standard DFW theory, which we will call the quadruplet representation of the Dirac theory (QRD), $\psi$ is defined to be a section of the trivial vector bundle $\mathrm{V} \times {\sf C}^4$. In the second representation of a wave function, which we will call the tensor representation of the Dirac theory (TRD),   $\psi$ is defined to be a section of the complex tangent bundle $\mathrm{T}_{\sf C}\mathrm{V}$.  Thus,  $\psi$ is either a quadruplet of four scalar fields for QRD or a four-vector field for TRD. The connection $D$  is a specific connection on the relevant vector bundle E in which the wave function $\psi$ is living, which is defined either as the spinor connection for DFW, or e.g. as the Levi-Civita connection extended to the complex tangent bundle $\mathrm{T}_{\sf C}\mathrm{V}$  for TRD \cite{A39}.  We will see in Section \ref{SpecialClasses} that many other choices for the connection $D$ are possible in both representations.\\  

\noindent Note that by the definition of a connection $D$ on a vector bundle E, it associates to any section $\psi$ of E, a section $D\psi$ of the tensor product bundle $\mathrm{T V^\circ \otimes E}$, where we denote the vector bundles dual to $\mathrm{T V}$ and E as $\mathrm{T V^\circ }$ and $\mathrm{E^\circ }$, respectively \cite{ChernChenLam1999}.\\  

Recall the following canonical isomorphisms of vector bundles (\cite{DieudonnéTome3}, Eq. (16.18.3.4)):
\be
Hom(\mathrm{E},\mathrm{F}) \cong \mathrm{F}\otimes \mathrm{E}^\circ \cong \mathrm{E}^\circ \otimes \mathrm{F},
\ee
where E and F are two vector bundles having in common the base space V.  Then, using the fact that smooth sections of the vector bundle $Hom(\mathrm{E},\mathrm{F})$ may be identified with smooth vector bundle maps $\mathrm{E} \rightarrow \mathrm{F}$, which project to the identity map on V (\cite{Husemoller}, p. 67), and finally, using the canonical isomorphism $\mathrm{T V} \cong \mathrm{TV}^\circ $ induced by the spacetime metric on the tangent bundle $\mathrm{T V}$, the Dirac matrices may be regarded as a smooth section $\gamma$ of the following vector bundle:
\bea\label{gamma bundle}\nonumber
\mathrm{TV}	 \otimes  \mathrm{E} \otimes \mathrm{E}^\circ & \cong & \mathrm{TV}^\circ	\otimes \mathrm{E} \otimes \mathrm{E}^\circ\\ 
& \cong & \mathrm{TV}^\circ	\otimes Hom(\mathrm{E},\mathrm{E}) \cong Hom\left(\mathrm{TV},Hom(\mathrm{E}, \mathrm{E})\right) 
\eea
Such a section will be called simply a ``$\gamma$ field".\\

\subsection{Local expressions}
 
Locally, by restricting to a sufficiently small open subset W of the spacetime V, we may select a frame field (or basis of vector fields) $(e_a)$ on the relevant vector bundle E.
\footnote{\ 
A notation like $(e_a)$ will designate an ordered family of elements indexed by a set of indices which is clear from the context. In this paper, the set of indices will always be $\{0,...,3\}$. 
}
 Restricting the wave function $\psi$ to W, the wave function may be expressed as:  
\be\label{psi=psi^a e_a}
\psi=\Psi^a\,e_a.
\ee
Then, choosing local coordinates $X\mapsto (x^\mu)$  in an open subset $\mathrm{U}\subset\mathrm{W}$, with the corresponding basis of coordinate vector fields $(\partial_\mu)\equiv \left(\frac{\partial}{\partial x^\mu}\right)$, which is a frame field on the tangent bundle TU, we have from Eq. (\ref{gamma bundle}): 
\be\label{gamma-intrinsic}
\gamma = \gamma^{\mu a } _b\ \partial_\mu \otimes e_a \otimes \theta^b,
\ee
where $(\gamma^{\mu a } _b)$ is the family of the complex coefficients of the tensor field $\gamma$, and $(\theta^a)$ is the frame field (or basis of one-forms) dual to the selected frame field $(e_a)$.  From the  field (\ref{gamma-intrinsic}), the Dirac matrices are defined to be the matrices with components:
\be\label{gamma^mu from gamma intrinsic}
\left(\gamma^\mu\right)^a _{\ \,b} \equiv \gamma^{\mu a } _{ b}. 
\ee
Thus, they are defined locally, and depend on the choice of local coordinates and local frame field.  These definitions give rise to the correct transformation behaviors.\\

Consider any section $\psi$ of the complex vector bundle E restricted to U. In the local frame $(\dd x ^\mu)$ on TU$^\circ$, dual of the coordinate frame $(\partial _\mu)$, and in the local frame field $(e_a)$ on E restricted to U, $D\psi$ has the local expression:
\be\label{Dpsi}
D\psi=D_\mu \Psi^b\ \dd x ^\mu \otimes e_b.
\ee
Accordingly, we have
\be\label{D_partial_mu psi}
D_{\partial_\mu} \psi \equiv (D\psi)(\partial_\mu)=D_\mu \Psi^b\ e_b.
\ee
In particular, we define the connection matrices $\Gamma_\mu$, whose components $\left(\Gamma_\mu\right)^b_{\ \,a}$ are determined from: 
\be\label{De_a}
D_{\partial_\mu} e_a=(\Gamma _\mu)^b_{\ \ a}\, e_b.
\ee
The components $D_\mu \Psi^b$ of $D\psi $ in Eq. (\ref{Dpsi}) can then be written by using Eqs. (\ref{psi=psi^a e_a}) and (\ref{De_a}):
\be\label{Dpsi-explicit}
D_\mu \Psi^b \equiv \frac{ \partial \Psi^b}{\partial x^\mu } + (\Gamma _\mu)^b_{\ \ a}\, \Psi^a.
\ee
In such local coordinates, the Dirac equation (\ref{Dirac slash}) reduces to the usual form: 
\be\label{Dirac-general}
\gamma ^\mu D_\mu\Psi = -i m \Psi,
\ee
where $\Psi$ is the column vector $(\Psi^a)$, and $D_\mu\Psi$ for each $\mu=0,... , 3$ is the column vector $(D_\mu \Psi^b)$.\\

For TRD, as in previous work \cite{A39,A40,A42,A43}, the frame field $(e_a)$ on the complex tangent bundle  $\mathrm{E=T}_{\sf C}\mathrm{V}$ can be taken to be the coordinate basis so that $e_a\equiv \delta^\mu_a \,\partial_\mu$.  In that case, the components of the wave function: $\Psi^\mu\equiv \Psi^a \,\delta^\mu_a$  transform as the components of a four-vector field after a coordinate change as follows:  Setting $L^\mu_{\ \nu} \equiv \frac{\partial x'^\mu }{\partial x^\nu }$, we have  for  the TRD wave function: 
\be\label{psi'-TRD-coordinate basis}
\Psi'^\mu=L^\mu_{\ \nu} \Psi^\nu.
\ee
Similarly, the components of the TRD Dirac matrices (\ref{gamma-intrinsic}) are then given by $\gamma^{\mu \rho} _\nu \equiv \gamma^{\mu a } _{ b}\,\delta^\rho_a \,\delta^b_\nu$ and transform as an affine $(^2 _1)$ tensor \cite{A39, A42}:
\be \label{gamma-(^2_1)tensor}
 \gamma'^{\mu \rho} _\nu = L^\mu_{\ \sigma }\,L^\rho _{\ \tau }\,\left(L^{-1}\right)^\chi_{\ \nu  }\gamma^{\sigma \tau } _\chi .
\ee
Whereas, the four scalar character of the wave function (\ref{psi=psi^a e_a}) for QRD means that we have the canonical basis of ${\sf C}^4$, namely:
\be\label{E_a}
E_0=(1,0,0,0),\quad E_1=(0,1,0,0),\quad E_2=(0,0,1,0),\quad E_3=(0,0,0,1)
\ee  
as a fixed frame field $(E_a)$ on the complex vector bundle $\mathrm{E}=\mathrm{V} \times {\sf C}^4$. Hence, the quadruplet of scalar fields $(\Psi^a)$ remains invariant during a coordinate change, and the Dirac matrices $\gamma^\mu$ in Eq. (\ref{gamma^mu from gamma intrinsic}) transform as a matrix-valued four-vector:
\be\label{gamma-vector}
\gamma'^\mu =L^\mu_{\ \nu}\, \gamma^\nu.
\ee

\noindent The anticommutation relation (\ref{Clifford}) is covariant under a change of chart, for either of the two transformation modes (\ref{gamma-vector}) and (\ref{gamma-(^2_1)tensor}) \cite{A37}.\\

In spinor theory (including DFW), tensor indices refer to three basic vector bundles and their duals.  The three basic vector bundles are the tangent bundle   TV, the spinor bundle E, and the complex conjugate spinor bundle $\mathrm{E}^*$.    Thus, there are four types of tensor indices:   middle and late Greek letters $\mu,\nu,...$  will be used for coordinate indices; early Greek letters $\alpha,\beta,...$ will be used for tetrad (or frame) indices for TV; early Latin letters $a,b,...$ will be used as frame indices for E; and $a^*,b^*,...$ will be used as frame indices for $\mathrm{E}^*$.  Note that contractions can only be performed for like indices. Finally, middle Latin letters $j , k ,...$ will be used as indices for spatial coordinates only ($ j , k = 1, 2, 3$). Throughout this paper  $(\eta ^{\alpha\beta })= (\eta _{\alpha\beta})=(\eta ^{ab })= (\eta _{ab})\equiv \mathrm{diag}(1,-1,-1,-1)$ will be used to denote Minkowski metrics.

\section{Special classes of Dirac equations}\label{SpecialClasses}

We will first introduce special classes of QRD equations and then TRD equations.  Within a given class, a continuum of different possibilities exist for the $\gamma$ field  \cite{A42,A43,B31}. The connection is fixed by the choice of the $\gamma$ field for DFW, but is chosen independently of the latter for the other four classes that we will introduce. Whichever class is chosen, the Dirac equation has either the normal form (\ref{Dirac-general}), or the modified (extended) form (\ref{Dirac-general-modified}) that we will introduce in Section \ref{modified Dirac equation}.

\subsection{The Dirac-Fock-Weyl (DFW) equation} 
This equation \cite{Weyl1929b,Fock1929b}, defined for sections of the trivial bundle $\mathrm{E}=\mathrm{V} \times {\sf C}^4$, is the standard form of the Dirac equation in a curved spacetime. It is a QRD equation characterized by two facts \cite{ChapmanLeiter1976}:\\

a) In any local coordinate domain $\mathrm{U} \subset \mathrm{V}$, the $\gamma$ field is expressed as a linear function of a fixed set of constant Dirac matrices, say $(\gamma ^{\natural \alpha })$, through a set of real coefficients $a^\mu_{\ \,\alpha}(X)$ varying with the spacetime point $X \in \mathrm{U}$: 
\be \label{flat-deformed}
  \gamma ^\mu(X) = a^\mu_{\ \,\alpha}(X)  \ \gamma ^{ \natural \alpha}.
\ee
Here the set $(\gamma ^{\natural \alpha })$ of ``flat" Dirac matrices is a constant solution of Eq. (\ref{Clifford}) above with the Minkowski metric $\eta ^{\alpha \beta }$ instead of the spacetime metric $g^{\mu\nu}$: 
\be \label{Clifford-flat}
\gamma ^{\natural \alpha } \gamma ^{\natural \beta  } + \gamma ^{\natural \beta  } \gamma ^{\natural \alpha } = 2\eta ^{\alpha \beta }\,{\bf 1}_4, \quad \alpha  ,\beta  \in \{0,...,3\}.
\ee
The coefficients $a^\mu_{\ \,\alpha}$ are the components with respect to the local coordinate basis on U of a real global orthonormal tetrad field $(u_\alpha)$ on V\,; i.e., a global orthonormal frame field on the tangent bundle TV. That is, $u_\alpha=a^\mu_{\ \,\alpha}\, \partial_\mu$. Therefore, the components $a^\mu_{\ \,\alpha}$ of $u_\alpha$ satisfy the orthonormality condition:
\be\label{orthonormal tetrad}
 g_{\mu \nu }\,a^\mu _{\ \,\alpha} \,a^\nu _{\ \,\beta } = \eta_{\alpha\beta},
\ee
[here $g_{\mu \nu } \equiv \,<\partial_\mu ,\partial_\nu >$], which ensures that the field of ``curved'' Dirac matrices $(\gamma^\mu)$ in Eq. (\ref{flat-deformed}) satisfies the anticommutation relation (\ref{Clifford}). \\

Globally, the DFW $\gamma$ field is given by:
\be\label{gamma-decompos-1}
\gamma = \gamma^{\alpha a}  _{b}\ u_\alpha \otimes E_a \otimes \Theta^b,
\ee
where $(E_a)$ is the canonical constant frame field (\ref{E_a}), and $(\Theta^a)$ denotes its dual frame field. Note that $E_a \otimes \Theta^b$ can be regarded as a matrix with one in the $a b$ position and zeros elsewhere.\\

\vspace{3mm}
b) The gamma field is {\it covariantly constant} with respect to the relevant connection, i.e., $D\gamma=0$.\\

The two conditions a) and b) lead to the form 
\be\label{Spin matrices}
\Gamma _\mu =\omega _{\mu \alpha \beta }\,s^{\alpha \beta}
\ee 
for the matrices of the connection $D$, called ``spin matrices'', with real coefficients $\omega _{\mu \alpha \beta  }$ and where $s^{\alpha \beta }\equiv [ \gamma ^{ \natural \alpha },\gamma ^{ \natural \beta }]$, and to determining the precise expression of the coefficients $\omega _{\mu \alpha \beta   }$ \cite{ChapmanLeiter1976}. It is found that this expression, and thus the corresponding ``spin connection'' $D$ itself, {\it depends on the field $\gamma$.} \\

In the literature, the DFW equation has been usually used with the following additional restriction on the set $(\gamma ^{\natural \alpha })$ of constant ``flat'' Dirac matrices:\\

c) The constant matrix $\gamma ^{\natural 0} $ is a hermitizing matrix \cite{Pauli1936,A40} for the Dirac matrices $\gamma^\mu(X)$ in Eq. (\ref{flat-deformed}). \\

\noindent This restriction, which usually is not explicitly stated (except for Refs. \cite{BrillWheeler1957+Corr,Audretsch1974p328}), is gotten by choosing a particular set $(\gamma ^{\natural \alpha })$, such that $\gamma ^{\natural 0} $ is in fact a hermitizing matrix for the set of constant ``flat'' Dirac matrices $(\gamma ^{\natural a })$ \cite{A42}.\\

\subsection{Other classes of Dirac equations}

We will now introduce four other interesting classes of Dirac equations: one is a QRD equation, the other three are TRD equations. For these four classes, we do not restrict the $\gamma$ field in any way beyond the necessity of satisfying the anticommmutation relation (\ref{Clifford}). Thus, each of these four classes is characterized by assuming a specific connection on the relevant vector bundle. The TRD--1 and TRD--2 equations were proposed in Ref. \cite{A39}. The QRD--0 and TRD--0 equations are new.

\subsubsection{The QRD--0 equation}\label{QRD-0}

We may introduce a very simple form of QRD equation by taking the {\it trivial connection} on the trivial bundle $\mathrm{E}=\mathrm{V} \times {\sf C}^4$:
\be\label{Trivial connection}
\Gamma _\mu =0 \qquad \mathrm{in\ the\ canonical\ frame\ field\ } (E_a).
\ee 
In view of (\ref{De_a}), this connection can be characterized by the fact that
\be\label{DE_a=0}
D\,E_a=0.
\ee

\vspace{5mm}

\subsubsection{The TRD--1 and TRD--2 equations}

For the TRD--1 equation, the connection is simply the {\it Levi-Civita connection} (extended to $\mathrm{T}_{\sf C}\mathrm{V}$). Thus, if the frame field $(e_a)$ on the complex tangent bundle $\mathrm{T}_{\sf C}\mathrm{V}$ is taken locally to be the coordinate basis $(\partial_\mu)$ associated with a chosen chart on V, and if the components $ (\Psi^a)$ of the four-vector wave function are taken locally as $ (\Psi^\mu)$, that is,  $e_a\equiv\delta^\mu_a \partial_\mu$ and $\Psi^\mu \equiv \delta^\mu_a \Psi^a$ (as was implicitly assumed in previous work \cite{A39,A42,A43}), then the connection coefficients are the Christoffel symbols of the second-kind associated with the spacetime metric $g_{\mu \nu }$:
\be\label{Levi-Civita}
\left(\Gamma_\mu\right)^\nu_{\ \,\rho}  \equiv  \left\{^\nu _{\rho \mu } \right\}.
\ee

\vspace{5mm}
For the other TRD equation (TRD--2), the connection is defined from the {\it spatial} Levi-Civita connection in an assumed {\it preferred reference frame} \cite{A39}. We will not need its explicit expression.

\subsubsection{The TRD--0 equation}\label{TRD-0}
For the TRD--0 equation, the connection $D$ is the so-called {\it teleparallel connection} associated with a given {\it orthonormal tetrad field} $(u_a)$ and its dual orthonormal tetrad field $(\omega^a)$ defined, respectively, on the complex tangent bundle $\mathrm{T}_{\sf C}\mathrm{V}$ and its dual $\mathrm{T}_{\sf C}\mathrm{V}^\circ$. That is, extending the spacetime metric $<\,,\,>$ to $\mathrm{T}_{\sf C}\mathrm{V}$ and $\mathrm{T}_{\sf C}\mathrm{V}^\circ$,  the orthonormal tetrad fields $(u_a)$ and $(\omega^a)$ satisfy:
\be
<u_a,u_b>\,=\eta_{ab}\quad;\quad <\omega^a,\omega^b>\,=\eta^{ab},
\ee
though here $(u_a)$ and $(\omega^a)$ are generally complex. The teleparallel connection is characterized by the fact that
\be\label{Du_a=0}
D\,u_a=0\quad;\quad D\,\omega^a=0. 
\ee
This connection depends on the tetrad field $(u_a)$, of course. Let us denote by $\mathcal{G}$ the metric tensor induced on $\mathrm{T}_{\sf C}\mathrm{V}$ by the spacetime metric as follows:  
\be\label{mathcal G}
\mathcal{G}= \eta _{ab }\, \omega^a \otimes\omega^b.
\ee
As a consequence of Eqs. (\ref{Du_a=0}) and (\ref{mathcal G}), the induced metric   is covariantly constant, $D\mathcal{G}=0$.

\vspace{5mm}

\section{Local similarity transformations}\label{LocalSimilarities}

The notion of a local similarity transformation (in short ``a local similarity'') has been recalled in the Introduction, as basically switching from one possible field of Dirac matrices to another one by Eq. (\ref{similarity-gamma}). Simultaneously, one usually transforms also the wave function according to Eq. (\ref{psitilde=S^-1 psi}). Whereas the latter equation occurs naturally in the case of a flat spacetime with affine coordinates \cite{A40}, in which the similarity transformation $S$ does not depend on the spacetime point $X$, it is less obvious in the general case. However, one possible definition of a local similarity is a mere change of the frame field $(e_a)$ on the vector bundle E, of which the wave function is a section:
\be\label{e-tilde}
\widetilde{e}_b=S^a_{\ \, b}\,e_a .
\ee
Under such a change, the column matrix $\Psi\equiv (\Psi^a)$, made with the components of the wave function (\ref{psi=psi^a e_a}) in the frame field, changes indeed according to Eq. (\ref{psitilde=S^-1 psi}), while the Dirac matrices $\gamma^\mu \equiv (\gamma^{\mu a } _{\quad b})$ associated with the $\gamma$ field (\ref{gamma-intrinsic}) change indeed according to Eq. (\ref{similarity-gamma})---the matrix $S$ in Eqs. (\ref{similarity-gamma})--(\ref{psitilde=S^-1 psi}) having components $S\equiv (S^a_{\ \, b})$. Since the form (\ref{similarity-gamma}) of the new matrices $\widetilde{\gamma} ^\mu$ ensures trivially that they satisfy the same anticommutation relation (\ref{Clifford}) as do the starting ones $\gamma ^\mu$, the relation (\ref{Clifford}) is thus covariant under a change of the frame field $(e_a)$ on E, as announced in Section \ref{Two representations}.\\

At the same time, it is easy to check that the connection matrices in Eq. (\ref{Dpsi-explicit}) change according to 
\be\label{Gamma-tilde-psitilde=S^-1 psi}
\widetilde{\Gamma }_\mu = S^{-1}\Gamma _\mu S+ S^{-1}(\partial _\mu S).
\ee
Under a local similarity, seen as a change (\ref{e-tilde}) of the frame field, the covariance of the Dirac equation (\ref{Dirac-general}) is an obvious fact: it is merely rewriting a tensor equation in another frame field. Thus, it applies to any version of the Dirac equation, of course. This kind of local similarity may be termed a ``passive'' one. Note that Eq. (\ref{Gamma-tilde-psitilde=S^-1 psi}) is the one stated by Chapman \& Leiter \cite{ChapmanLeiter1976} to ensure that the DFW equation remains covariant after what they call a ``spin transformation'', which designates indeed a local similarity. Clearly, Eq. (\ref{Gamma-tilde-psitilde=S^-1 psi}) applies to any connection on a vector bundle after a change (\ref{e-tilde}) of the frame field. \\

On the other hand, one may also consider ``active'' local similarities. Then, one leaves  the frame field $(e_a)$ unchanged, and one defines a new gamma field $\widetilde{\gamma}$ and a new wave function $\widetilde{\psi}$, whose local expressions in the fixed frame field $(e_a)$ are related to the local expressions of $\gamma $ and $\psi $ by the same Eqs. (\ref{similarity-gamma})--(\ref{psitilde=S^-1 psi}) as for a passive similarity. Thus, the relations between the components of the wave functions, $\Psi$ and $\widetilde{\Psi}$, and between the matrices $\gamma^\mu$ and $\widetilde{\gamma}^\mu$, are the same Eqs. (\ref{similarity-gamma})--(\ref{psitilde=S^-1 psi}) as for a passive similarity. It follows that, also for an active similarity, the Dirac equation (\ref{Dirac-general}) is covariant iff one changes the connection matrices in Eq. (\ref{Dpsi-explicit}) according to Eq. (\ref{Gamma-tilde-psitilde=S^-1 psi}).\\ 

For the DFW equation, the local similarities are restricted to the spin group, as mentioned in the Introduction: $\forall X \in \mathrm{V},\ S(X) \in {\sf Spin(1,3)}$. This is due to the fact that the $\gamma$ field is defined from an orthonormal tetrad field, Eq. (\ref{flat-deformed}). In Section \ref{Relations}, we will study some correspondences between different classes of Dirac equations introduced in Section \ref{SpecialClasses}. Therefore, the local similarity matrices $S(X)$ can be any element of the linear group ${\sf GL(4, C)}$.\\

Note that, from the fact that it transforms a frame field on E into another one, it follows that a local similarity, either passive or active, is (associated with) a section of the vector bundle $\mathrm{E} \otimes \mathrm{E}^\circ$:
\be
\mathcal{S} \equiv S^a_{\ \, b}\, e_a \otimes \theta^b.
\ee
The matrix $S$ is thus the matrix of the components of $\mathcal{S}$ in the local frame field $(e_a)$, with dual frame field $(\theta ^a)$. The connection induced on $\mathrm{E} \otimes \mathrm{E}^\circ$ by the connection $D$ on E allows us to define the covariant derivatives of $\mathcal{S}$:
\be
D_\mu S^a_{\ \, b}=\partial_\mu S^a_{\ \, b} + (\Gamma _\mu)^a_{\ \ c}\,S^c_{\ \, b} - (\Gamma _\mu)^c_{\ \ b}\,S^a_{\ \, c},
\ee
or in matrix form:
\be\label{D_mu S}
D_\mu S \equiv \left( D_\mu S^a_{\ \, b}\right)=\partial_\mu S + \Gamma_\mu\,S - S\,\Gamma_\mu.
\ee

\vspace{5mm}

\section{The modified Dirac equation}\label{modified Dirac equation}

The Lagrangian (\ref{Lagrangian-intrinsic}) extends the standard Dirac Lagrangian valid for the DFW equation (e.g. \cite{BrillWheeler1957+Corr, Leclerc2006}), in that it is valid in general for QRD and TRD theory, and it involves the hermitizing metric $(\,,\,)$, introduced by Pauli \cite{Pauli1936,Pauli1933}. The latter can be regarded as a tensor field $\mathcal{A}$, more precisely as a section of the vector bundle $(\mathrm{E}^\circ)^* \otimes \mathrm{E}^\circ$. Its local expression in a given frame field $(e_a)$, with dual frame field $(\theta^a)$, is:
\be\label{A-local expression}
\mathcal{A} = A_{a^*\, b}\, \theta^{*\, a^*} \otimes\theta^b,
\ee
where we employ the canonical conjugate linear isomorphism $\mathrm{E}^\circ \rightarrow (\mathrm{E}^\circ)^*$, which, at every spacetime point $X \in \mathrm{V}$, maps each covector $\theta \in \mathrm{E}_X^\circ$ to the conjugate covector $\theta^* \in (\mathrm{E}_X^\circ)^*$ defined by $\theta^*(u)=\theta(u)^*$ for all $u \in \mathrm{E}_X$.  Here $\theta(u)^*$ denotes the ordinary complex conjugate of the complex number $\theta(u)$. We note that Eq. (\ref{A-local expression}) ensures that the hermitizing matrix $A\equiv(A_{a^*b})$ satisfies the correct transformation behaviour under a local similarity, i.e., under a change (\ref{e-tilde}) of the frame field, namely \cite{A40}:
\be \label{similarity-A}
\widetilde{A}=  S^\dagger  A S.
\ee
Previous work \cite{A40,A42} has proved the existence and uniqueness, up to a real factor $\lambda(X) \ne 0$, of the hermitizing matrix field $A=A(X)$, in any spacetime. 

\vspace{5mm}
The local expression of the Lagrangian density associated with the Lagrangian (\ref{Lagrangian-intrinsic}) is thus given, using Eqs. (\ref{Dpsi-explicit}) and (\ref{Dirac-general}), by:
\be\label{Lagrangian-density}
l=\sqrt{-g}\ \frac{i}{2}\left[\,\overline{\Psi}\gamma^{\mu}(D_{\mu}\Psi)-
\left(\overline{D_{\mu}\Psi} \right)\gamma^{\mu}\Psi+2im\overline{\Psi}\Psi\right],
\ee
where $\overline{\Psi}\equiv \Psi ^\dagger A$, with $ \Psi ^\dagger$ denoting the complex conjugate transpose of $\Psi$, and similarly $\overline{D_{\mu}\Psi}\equiv \left(D_{\mu}\Psi \right) ^\dagger A$. In the general case that we are considering, it is straightforward to check that the Euler-Lagrange equation of this Lagrangian density gives the following general Dirac equation:
\be\label{Dirac-general-modified}
\gamma ^\mu D_\mu\Psi=-im\Psi-\frac{1}{2}A^{-1}(D_\mu B^\mu )\Psi,\\
\ee
where $B^\mu \equiv A \gamma^\mu$. Eq. (\ref{Dirac-general-modified}) was found from a different route in Ref. \cite{A42}, and previously called a {\it modified Dirac equation.}  Note that the general Dirac equation (\ref{Dirac-general-modified}) derived from the Lagrangian density (\ref{Lagrangian-density}) coincides with the normal Dirac equation (\ref{Dirac-general}), iff 
\be\label{D_mu B^mu=0}
D_\mu B^\mu=0,
\ee 
which was shown to be a special condition that the coefficient fields $(\gamma,\mathcal{A})$ of any normal Dirac equation (\ref{Dirac-general}) must satisfy in order to conserve the probability current \cite{A42}.\\

Eq. (\ref{D_mu B^mu=0}) is of course satisfied in DFW, since the coefficient fields $(\gamma,\mathcal{A})$ are covariantly constant. Thus, the Dirac equation in DFW is always normal.  For normal QRD and TRD equations, we do not require that the coefficient fields $(\gamma,\mathcal{A})$ be covariantly constant. As we will see in Section \ref{Relations}, normal Dirac equations exist locally for any connection $D$ on the complex vector bundle E.    
\\

Let us compute explicitly the following expressions involving covariant derivatives, as function of the connection matrices (\ref{De_a}). First, we may rewrite Eq. (\ref{Dpsi-explicit})
as 
\be\label{Dpsi-Gamma}
D_\mu \Psi =\partial_\mu\Psi + \Gamma _\mu \Psi.
\ee 
We have (Eqs. (33) and (35) of Ref. \cite{A42}):
\be\label{D_mu gamma^nu}
D_\mu \gamma ^\nu \equiv \partial _\mu \gamma ^\nu + \left\{^\nu _{\rho \mu } \right\}\gamma ^\rho + \left[\Gamma _\mu,\gamma ^\nu \right],
\ee
(where $[M,N]\equiv MN-NM$), and
\be\label{D_mu A}
D_\mu A \equiv \partial _\mu A - A\Gamma _\mu - \Gamma _\mu^\dagger A.
\ee
From this, it follows by Leibniz' rule [and since $B^\mu =A\gamma ^\mu$\ ]:
\be\label{D_mu B^nu}
D_\mu B^\nu=\partial_\mu B^\nu + \left\{^\nu _{\rho \mu } \right\}B ^\rho -B^\nu\Gamma_\mu-\Gamma_\mu^\dagger B^\nu.
\ee

\vspace{3mm}

\section{Relations between different classes of Dirac equations}\label{Relations}

In this section, we will prove that local similarity transformations transform the standard DFW equation into a particular linear TRD--1 equation which lives on the complex tangent bundle $\mathrm{T}_{\sf C}\mathrm{V}$, which inherits the Levi-Civita connection.  We achieve this by combining two maps: QRD $\rightarrow$ TRD  $\rightarrow$ TRD--1. We will need the following theorem of linear hyperbolic partial differential equations:\\

\paragraph{Theorem 0}\label{Theorem0} (Lax \cite{Lax2006}). {\it Let $M_1$, $M_2$,...,$M_n$ and $F$ be real $d \times d$  matrix functions that depend smoothly on $n+1$ independent real variables $t,x_1,x_2,...,x_n$ in a slab $-T\leq t\leq T, \ \ x\in {\sf R}^n$, denoted as $\mathrm{I}\times {\sf R}^n$.  Furthermore, suppose that $M_1$, $M_2$,...,$M_n$ are symmetric matrices.  Then the real linear hyperbolic system:
\be\label{LaxSystem}
\frac{\partial v}{\partial t} + \sum_{j=1}^n {M_j \frac{\partial v}{\partial x_j}}=Fv
\ee
has a smooth real vector valued solution $v : \mathrm{I}\times {\sf R}^n \rightarrow {\sf R}^d$ satisfying prescribed smooth initial data at $t=0$.}\\

\vspace{5mm}
\noindent Theorem 0 extends to several corollaries. First, a more symmetric form of Theorem 0 is given by:

\paragraph{Corollary 1.}\label{Corollary1} {\it Theorem 0 extends to the real linear hyperbolic system: 
\be\label{GaraSystem}
M_0\frac{\partial v}{\partial t} + \sum_{j=1}^n {M_j \frac{\partial v}{\partial x_j}}=Fv,
\ee
where $M_0$ is a smooth positive definite real $d\times d$ matrix function.}\\

\noindent {\it Proof}  (\cite{Garabedian1964}). Since $M_0$ is smooth and positive definite, it has a smooth Cholesky factorization $M_0=C^T\,C$, where $C^T$ denotes the transpose of the non-singular real matrix function $C$.  Then, substituting $v=C^{-1}w$ into Eq. (\ref{GaraSystem}) reduces it to the same form as in Eq. (\ref{LaxSystem}).  Q.E.D.\\

\vspace{5mm}
\noindent Next, Theorem 0 can be extended to complex equations as follows, by considering their real and imaginary parts:\\

\paragraph{Corollary 2.}\label{Corollary2} {\it  Let $M_0$, $M_1$,...,$M_n$ and $F$ be complex $d\times d$ matrix functions that depend smoothly on $n+1$ independent real variables  $t,x_1,x_2,...,x_n$ in a slab $-T\leq t\leq T, \ \ x\in {\sf R}^n$, denoted as $\mathrm{I}\times {\sf R}^n$.  Furthermore, suppose that $M_0$, $M_1$,...,$M_n$ are Hermitian matrices and $M_0$ is positive definite.  Then, the complex linear hyperbolic system:
\be\label{GaraSystem-C}
M_0\frac{\partial v}{\partial t} + \sum_{j=1}^n {M_j \frac{\partial v}{\partial x_j}}=Fv
\ee
has a smooth complex vector valued solution $v : \mathrm{I}\times {\sf R}^n \rightarrow {\sf C}^d$ satisfying prescribed smooth initial data at $t=0$.}.\\

\vspace{5mm}
\noindent Finally, in Section \ref{Dirac to QRD--0 or TRD--0} we will need the following matrix form of Theorem 0:\\

\paragraph{Corollary 3.}\label{Corollary3} {\it  Let $M_0$, $M_1$,...,$M_n$ and $F$ be as in Corollary 2, except that $F$ is now a homogeneous linear function of $d\times d$  matrices, as well as having explicit dependence on $t,x_1,x_2,...,x_n$.  Then the complex linear  hyperbolic system:
\be\label{GaraSystem-Matrix}
M_0\frac{\partial S}{\partial t} + \sum_{j=1}^n {M_j \frac{\partial S}{\partial x_j}}=F(S)
\ee
has a smooth complex matrix valued solution $S : \mathrm{I}\times {\sf R}^n \rightarrow \mathrm{M}({\sf C},d)$ which equals the identity matrix at $t=0$, as its prescribed smooth initial data.}\\

\noindent {\it Proof.}  Define a column vector $v : \mathrm{I}\times {\sf R}^n \rightarrow {\sf C}^{d^2}$ made from the successive columns of the matrix solution $S$:
\be\label{v}
v=\left(S^1_{\ \ 1},S^2_{\ \ 1},...,S^d_{\ \ 1},.....,S^1_{\ \ d},S^2_{\ \ d},...,S^d_{\ \ d}\right)^T.
\ee
Then, $M_0$, $M_1$,...,$M_n$ acting on the $d^2$ components of $v$ are embedded into $d^2\times d^2$ block diagonal matrices, which have the same Hermitian and positive definite properties that $M_0$, $M_1$,...,$M_n$ have in Corollary 2.  Also, note that $F$ acts linearly on the  $d^2$ components of $v$ in Eq. (\ref{v}), as in Eq. (\ref{GaraSystem-C}) of \hyperref[Corollary2]{Corollary 2}.  It is straightforward then to show that Eq. (\ref{GaraSystem-Matrix}), expressed in terms of the vector valued solution $v : \mathrm{I}\times {\sf R}^n \rightarrow {\sf C}^{d^2}$, reduces to the form of Eq. (\ref{GaraSystem-C}).  Q.E.D.

\subsection{Correspondence between the QRD and TRD equations}

\paragraph{Theorem 1.}\label{Theorem1} {\it In a non-compact, four-dimensional spacetime V that admits a spinor structure, any form of the QRD equation defined on the trivial bundle $\mathrm{V} \times {\sf C}^4$ is equivalent to a TRD equation defined on the complex tangent bundle $\mathrm{T}_{\sf C}\mathrm{V}$.  Moreover, any normal QRD equation is equivalent to a normal TRD equation.}\\

\noindent {\it Proof.} Consider the general Dirac equation (\ref{Dirac-general-modified}) in the QRD version, with an arbitrary connection $D$ on $\mathrm{V} \times {\sf C}^4$, and with an arbitrary $\gamma$ field. The latter is such that the Dirac matrices $(\gamma ^\mu )$ associated with it by Eq. (\ref{gamma^mu from gamma intrinsic}) on the domain $\mathrm{U} \subset\mathrm{V}$ of each coordinate chart satisfy the anticommutation relation (\ref{Clifford}). As recalled in Section \ref{Two representations}, in a non-compact, four-dimensional spacetime V that admits a spinor structure, there exists a global tetrad field $(u_\alpha)$. Thus, $(u_\alpha)$ is a global frame field on the tangent bundle TV. Also, there is a constant canonical frame field $(E_a)$ on the trivial bundle $\mathrm{V} \times {\sf C}^4$, with the corresponding dual frame field $(\Theta^a)$. Hence, the $\gamma$ field has a unique global expression:
\be\label{gamma-decompos}
\gamma = \gamma^{\alpha a}  _{b}\ u_\alpha \otimes E_a \otimes \Theta^b.
\ee
The global frame field $(u_\alpha )$ on the tangent bundle TV induces a global frame field $(u_a)$ on the complex tangent bundle $\mathrm{T}_{\sf C}\mathrm{V}$, where $u_a=\delta ^\alpha _a u_\alpha $.  Thus, there is at least one global frame field on $\mathrm{T}_{\sf C}\mathrm{V}$, showing that it too is parallelizable.  Now, let $(e_a)$ be {\it any} fixed global frame field on the parallelizable complex tangent bundle $\mathrm{T}_{\sf C}\mathrm{V}$.  Then, we may associate with $\gamma $ a gamma field $\gamma '$ relevant to TRD, by setting
\be\label{gamma'}
\gamma' = \gamma^{\alpha a}  _{b}\ u_\alpha \otimes e_a \otimes \theta^b,
\ee 
where $(\theta^a)$ is the dual frame field of $(e_a)$. That is, the field $\gamma'$ has the same components as $\gamma$. It follows easily from Eqs. (\ref{gamma-intrinsic}) and (\ref{gamma^mu from gamma intrinsic}) that the matrices $\gamma'^\mu$ associated with $\gamma'$ on each coordinate domain $\mathrm{U} \subset\mathrm{V}$ have the same components as the matrices $\gamma^\mu$ associated with $\gamma$. Hence, both $\gamma^\mu$ and $\gamma'^\mu$ satisfy the anticommutation relation (\ref{Clifford}).\\

In the same way, we associate to the Hermitian metric $\mathcal{A}$ that is hermitizing for the $\gamma$ matrices, a Hermitian metric $\mathcal{A}'$ that is hermitizing for the $\gamma'$ matrices:
\be\label{A'}
A_{a^*\, b}\, \Theta^{*\, a^*} \otimes\Theta^b =\mathcal{A} \mapsto \mathcal{A}'=A_{a^*\, b}\, \theta^{*\, a^*} \otimes\theta^b.
\ee

\vspace{1mm}
\noindent Indeed, the map $(E_a)\mapsto (e_a)$ between the two frame fields induces a vector bundle isomorphism $\mathrm{V} \times {\sf C}^4 \rightarrow \mathrm{T}_{\sf C}\mathrm{V}$, so that we associate to any section $\psi$ of $\mathrm{V} \times {\sf C}^4$, a section $\psi'$ of $\mathrm{T}_{\sf C}\mathrm{V}$ (and conversely), as follows:
\be\label{psi'}
\Psi^a E_a=\psi \mapsto \psi'=\Psi^a e_a.
\ee
The isomorphism (\ref{psi'}) associates with any connection $D$ on $\mathrm{V} \times {\sf C}^4$, a connection $D'$ on $\mathrm{T}_{\sf C}\mathrm{V}$ (and conversely). Specifically, the coefficients of the connections $D$ and $D'$, with respect to the corresponding frame fields $(u_\alpha,E_a)$ and $(u_\alpha,e_a)$, are set equal to each other in this association. Moreover, as discussed above, the anticommutation relation of the $\gamma $ field and the hermitizing property of the Hermitian metric $\mathcal{A}$ are preserved by the isomorphism. It follows then that the global expression of the general Dirac equation (\ref{Dirac-general-modified}) is identical for the components of a QRD equation and its associated TRD equation with respect to the corresponding global frame fields $(u_\alpha,E_a)$ and $(u_\alpha,e_a)$. In the same way, the global expression of the normal condition (\ref{D_mu B^mu=0}) is identical for the components of a QRD equation and its associated TRD equation with respect to the corresponding global frame fields $(u_\alpha,E_a)$ and $(u_\alpha,e_a)$. Therefore, the normal condition (\ref{D_mu B^mu=0}) is also preserved. Q.E.D.

\vspace{3mm}


\subsection{Transforming a Dirac equation to a QRD--0 or a TRD--1 equation}\label{Dirac to QRD--0 or TRD--0}


\paragraph{Theorem 2.}\label{Theorem2}  {\it Consider any form of the QRD (or TRD) equation with connection $D$ defined on the complex vector bundle $\mathrm{E}$ ($\mathrm{E=V} \times {\sf C}^4$ for QRD, $\mathrm{E=T}_{\sf C}\mathrm{V}$ for TRD). Let $D'$ be any other connection on $\mathrm{E}$.  Let $\chi:\ \mathrm{V}\supset\mathrm{U} \rightarrow {\sf R}^4$ be any chart of the spacetime $\mathrm{V}$ such that $\chi(\mathrm{U}) \supset \mathrm{I}\times {\sf R}^3$. Suppose that the spacetime metric $g_{\mu\nu}$ in $\mathrm{U}$ satisfies $g_{00}>0$ and the $3\times3$ matrix $(g_{jk})$ is negative definite. Then, there exists a local similarity transformation $S$, defined in an open domain $\mathrm{W} \subset \mathrm{U}$ satisfying $\chi(\mathrm{W}) \supset \{0\}\times {\sf R}^3$, which transforms the QRD (or TRD) equation, restricted to the domain $\mathrm{W}$, into a QRD (or TRD) equation with connection $D'$.  Furthermore, the local similarity transformation $S$ transforms any normal QRD (or TRD) equation into a normal QRD (or TRD) equation on the domain $\mathrm{W}$.}\\

\vspace{5mm}
\noindent {\it Proof.} In the domain of each chart: $\mathrm{U} \subset\mathrm{V}$, with coordinates $(x^\mu)$, the Dirac operator entering the general Dirac equation (\ref{Dirac-general-modified}) is given by:
\be\label{Dirac-modified-operator}
\mathcal{D}\equiv \gamma ^\mu D_\mu +\frac{1}{2}A^{-1}(D_\mu B^\mu ).
\ee
Since the Christoffel symbols in Eq. (\ref{D_mu B^nu}) satisfy $\left\{^\mu _{\rho \mu } \right\}=[\partial _\rho (\sqrt{-g})]/\sqrt{-g}$, we have on contracting the index $\mu$:
\be\label{D_mu A gamma^mu}
D_\mu B^\mu=\frac{1}{\sqrt{-g}}\partial _\mu \left(\sqrt{-g}A\gamma ^\mu \right)-A\Gamma-\Gamma^\dagger A ,
\ee
where $\Gamma$ is the matrix
\be
\Gamma \equiv \gamma ^\mu \Gamma _\mu.
\ee

\vspace{3mm}
Now let $D$ and $D'$ denote any two connections on E. Let $\Gamma _\mu$ and $\Gamma' _\mu $ denote the connection matrices for the two connections $D$ and $D'$. Suppose the coefficient fields $(\gamma ^\mu,A)$ are the same for both connections $D$ and $D'$, and let $\mathcal{D}$ and $\mathcal{D}'$ denote the respective Dirac operators. Set $K_\mu =\Gamma _\mu -\Gamma' _\mu$. We get from Eq. (\ref{D_mu A gamma^mu})
\be\label{D_mu A gamma^mu vs TRD--1}
D_\mu B^\mu=D' _\mu B^\mu-AK-K^\dagger A ,
\ee
where
\be\label{K}
K \equiv \gamma ^\mu K _\mu.
\ee
It follows from Eqs. (\ref{Dirac-modified-operator}) and (\ref{D_mu A gamma^mu vs TRD--1}) that 
\be\label{Dirac operator vs TRD--1}
\mathcal{D}=\mathcal{D}'+\frac{1}{2}A^{-1}\left( AK-K^\dagger A \right),
\ee
provided that the coefficient fields $(\gamma ^\mu,A)$ are the same for both Dirac operators $\mathcal{D}$ and $\mathcal{D}'$.\\

More specifically, let $D$ and $D'$ be the two connections mentioned in the
theorem. Consider then a local similarity transformation $S$, defined on a
domain W as described in the theorem, mapping the connection $D$ with
connection matrices $\Gamma _\mu $ to a new connection $\widetilde{D}$ with connection matrices
$\widetilde{\Gamma} _\mu$, according to Eq. (\ref{Gamma-tilde-psitilde=S^-1 psi}), and mapping the coefficient fields $(\gamma ^\mu,A)$ of the starting Dirac equation to new coefficient fields $(\widetilde{\gamma} ^\mu ,\widetilde{A})$ according to Eqs. (\ref{similarity-gamma}) and (\ref{similarity-A}). Let the coefficient fields $(\widetilde{\gamma} ^\mu ,\widetilde{A})$ be the same for both connections $\widetilde{D}$ and $D'$. Let $\widetilde{\mathcal{D}}$ and $\mathcal{D}'$ denote the respective Dirac operators (\ref{Dirac-modified-operator}). As in Eq. (\ref{K}), define $K \equiv \gamma ^\mu K _\mu$, where $K_\mu =\Gamma _\mu -\Gamma' _\mu$, with $\Gamma' _\mu$ the connection matrices for the connection $D'$. Similarly, define $\widetilde{K}  \equiv  \widetilde{\gamma}^\mu \widetilde{K}_\mu$, where $\widetilde{K}_\mu= \widetilde{\Gamma} _\mu - \Gamma' _\mu $. Since the coefficient fields $(\widetilde{\gamma} ^\mu ,\widetilde{A})$ are the same for both Dirac operators $\widetilde{\mathcal{D}}$ and $\mathcal{D}'$, Eq. (\ref{Dirac operator vs TRD--1}) applies in the form
\be\label{Dirac operator vs TRD--1-tilde}
\widetilde{\mathcal{D}}=\mathcal{D}'+\frac{1}{2}\widetilde{A}^{-1}\left( \widetilde{A}\widetilde{K}-\widetilde{K}^\dagger \widetilde{A} \right).
\ee
Suppose that we can determine $S$, defined on the domain W as described in
the theorem, in such a way that $\widetilde{K}=0$. Then, from Eq. (\ref{Dirac operator vs TRD--1-tilde}) we have $\widetilde{\mathcal{D}}=\mathcal{D}'$, and hence the Dirac equation (\ref{Dirac-general-modified}) based on the connection $D'$  and the coefficient fields $(\widetilde{\gamma} ^\mu ,\widetilde{A})$ will be equal to the Dirac equation based on the connection $\widetilde{D}$ and the same coefficient fields $(\widetilde{\gamma} ^\mu ,\widetilde{A})$. In turn, the latter Dirac equation is equivalent via the local similarity transformation $S$ to the Dirac equation based on the connection $D$ and the coefficient fields $(\gamma ^\mu,A)$ [since, as stated above, the local similarity transformation $S$ changes the connection $D$ and the connection matrices $\Gamma _\mu $ to $\widetilde{D}$ and $\widetilde{\Gamma} _\mu$, respectively, according to Eq. (\ref{Gamma-tilde-psitilde=S^-1 psi}) --- see Section \ref{LocalSimilarities}]. Thus, any Dirac equation based on the connection $D$ will have been shown to be equivalent, in a domain W as described in the theorem, to a Dirac equation based on the connection $D'$, hence proving the first assertion of the theorem. \\

Thus, it remains to determine $S$, defined on the domain W as described in
the theorem, in such a way that $\widetilde{K} =0$. Substituting $\Gamma _\mu=\Gamma' _\mu +K_\mu$  and $\widetilde{\Gamma} _\mu=\Gamma' _\mu +\widetilde{K}_\mu$  into Eq. (\ref{Gamma-tilde-psitilde=S^-1 psi}), 
we have:
\be
\Gamma' _\mu +\widetilde{K}_\mu =S^{-1}(\Gamma' _\mu +K_\mu ) S + S^{-1}\partial _\mu S,
\ee
from which, using Eq. (\ref{D_mu S}), we obtain:
\bea\nonumber
\widetilde{K}_\mu & = & S^{-1}K_\mu  S + S^{-1}\left (\partial _\mu S + \Gamma' _\mu S - S \Gamma' _\mu\right )\\
& = & S^{-1}K_\mu  S + S^{-1} D' _\mu S.
\eea
Thus from Eqs. (\ref{similarity-gamma}) and (\ref{K}):
\bea\label{tildeK}\nonumber
\widetilde{K} & \equiv & \widetilde{\gamma}^\mu \widetilde{K}_\mu = S^{-1}\gamma ^\mu K_\mu S+S^{-1}\gamma ^\mu D' _\mu S\\
& = & S^{-1}K  S + S^{-1}\gamma ^\mu D' _\mu S.
\eea
Setting $\widetilde{K} =0$ in Eq. (\ref{tildeK}) and multiplying the resulting equation by $AS$, and since $B^\mu\equiv A\gamma^\mu$, we get:
\be\label{PDE S}
B ^\mu D' _\mu S =-AKS.
\ee      
Now, the matrix valued functions $B^\mu$ are Hermitian and $B^0$  is positive definite [10]
.  Thus, Eq. (\ref{PDE S}) is of the form of Eq. (\ref{GaraSystem-Matrix}). The existence of a smooth solution $S$ to Eq. (\ref{PDE S}) in the open domain $\mathrm{U}'\equiv \chi^{-1}(]-T,+T[\times {\sf R}^3 ) \subset \mathrm{U}$, with $S(X)$ being equal to the identity matrix ${\bf 1}_4$ when $t\equiv \mathrm{Proj}_1(\chi(X))=0$, hence follows from the hypothesis $\chi(\mathrm{U}) \supset [-T,+T]\times {\sf R}^3 $ and from \hyperref[Corollary3]{Corollary 3}. Denote by W the open subset of U$'$, hence of V, in which $\mathrm{det}(S)\ne 0$. Thus both $S$ and $S^{-1}$ are smooth matrix valued functions defined on the open domain $\mathrm{W}$. Since $S(X)={\bf 1}_4$ when $t\equiv \mathrm{Proj}_1(\chi(X))=0$, it follows that $\chi(\mathrm{W}) \supset \{0\}\times {\sf R}^3$.\\  

Finally, let us check the preservation of the normal form of the Dirac equation. After the local similarity transformation $S$, for which the connection matrices change according to Eq. (\ref{Gamma-tilde-psitilde=S^-1 psi}), a straightforward evaluation shows that $\widetilde{D}_\mu \widetilde{B }^\mu =S^\dagger (D_\mu B ^\mu ) S$, where $\widetilde{B }^\mu=\widetilde{A }\widetilde{\gamma }^\mu$. Moreover, from Eq. (\ref{D_mu A gamma^mu vs TRD--1}), since $\widetilde{K} =0$, we have 
\be\label{D_mu A gamma^mu vs TRD--1-tilde}
S^\dagger (D_\mu B ^\mu ) S= \widetilde{D} _\mu \widetilde{B}^\mu = D'_\mu \widetilde{B}^\mu - \widetilde{A}\widetilde{K}-\widetilde{K}^\dagger \widetilde{A}=D'_\mu \widetilde{B}^\mu.
\ee
Thus, if $D_\mu B ^\mu=0$, then $D'_\mu \widetilde{B }^\mu =0$. Therefore, the normal form (\ref{Dirac-general}) of the Dirac equation, when it occurs, is preserved. Q.E.D. \\

\paragraph{Theorem 3.}\label{Theorem3} {\it Consider any DFW equation defined on the spacetime $\mathrm{V}$, and let $\chi: \mathrm{V} \supset \mathrm{U}\rightarrow {\sf R}^4$ be any chart of the space time $\mathrm{V}$, such that $\chi(\mathrm{U}) \supset \mathrm{I}\times {\sf R}^3$, for which the spacetime metric $g_{\mu\nu}$ in $\mathrm{U}$ satisfies $g_{00}>0$ and the $3\times3$ matrix $(g_{jk})$ is negative definite.  Then, there is an open domain $\mathrm{W} \subset \mathrm{U}$ satisfying $\chi(\mathrm{W}) \supset \{0\} \times {\sf R}^3$, such that the DFW equation is equivalent to a normal TRD--1 equation on the domain $\mathrm{W}$. } \\

\vspace{5mm}
\noindent {\it Proof.}  By \hyperref[Theorem1]{Theorem 1}, any DFW equation, being a normal QRD equation, is equivalent to some normal TRD equation, on the whole of the spacetime. The assumptions of Theorem 3 then allow us to apply \hyperref[Theorem2]{Theorem 2} to state that the latter normal TRD equation is equivalent to a normal TRD--1 equation on some open domain $\mathrm{W} \subset \mathrm{U}$ satisfying $\chi(\mathrm{W}) \supset \{0\} \times {\sf R}^3$. Therefore, the starting DFW equation is equivalent to a normal TRD--1 equation on the domain $\mathrm{W}$.  Q.E.D. \\

Note that the metric conditions in Theorems 2 and 3, namely, $g_{00}>0$ and the $3\times3$ matrix $(g_{jk})$ is negative definite, are satisfied by almost all spacetime metrics $g_{\mu \nu }$ of interest. As shown in previous work, these metric conditions guarantee the existence of a Hilbert space with a positive definite scalar product for every Dirac equation \cite{A42}. A notable exception is the G\"odel spacetime, for which there exist no complete three-dimensional submanifolds which are space-like, and for which a positive definite Hilbert space scalar product cannot be defined \cite{A44}.

\section{Conclusion}\label{Conclusion}  

In a curved spacetime, there are only two ways for defining the Dirac wave function $\psi$ describing spin-half particles: First, it can be defined as a quadruplet of complex scalar fields. This is the quadruplet representation of the Dirac field (QRD), to which the standard Dirac equation (DFW) in a curved spacetime belongs \cite{Weyl1929b}-\cite{ChapmanLeiter1976}. Or, $\psi $ can be defined as a complex four-vector field. This is the tensor representation of the Dirac field (TRD), to which belong two alternative versions of the Dirac equation in a curved spacetime called TRD--1 and TRD--2, which were proposed recently \cite{A39,A42}.\\

We first presented these two different representations (QRD and TRD) in a common geometrical framework that includes common intrinsic definitions for the wave functions $\psi$, the coefficient fields $(\gamma,\mathcal{A})$, the connections $D$, as well as the Lagrangians from which the equations are derived. In this framework, we introduced two simple forms of the Dirac equation, namely, the \hyperref[QRD-0]{QRD--0} and \hyperref[TRD-0]{TRD--0} versions, in which the connection matrices are zero in a chosen frame field.  We then proved that the two representations (QRD and TRD)  are equivalent for corresponding wave functions $\psi$, coefficient fields $(\gamma,\mathcal{A})$, and connections $D$  (\hyperref[Theorem1]{Theorem 1}).\\

As a consequence of \hyperref[Theorem2]{Theorem 2}, any form of the QRD equation is equivalent to a QRD--0 equation, and any form of the TRD equation is equivalent to a TRD--1 equation, in the same spacetime.  From \hyperref[Theorem1]{Theorem 1} and \hyperref[Theorem2]{Theorem 2}, we may conclude more generally the following: The Dirac equation, either in the QRD or the TRD representation, can be written with any connection on the corresponding vector bundle, but any specific choice of the connection within either of the two representations can account for a variety of linear, covariant Dirac equations in a curved spacetime, that reduce to the original Dirac equation in a Minkowski spacetime. This does not mean that any two Dirac equations on a given curved spacetime are equivalent, which indeed is not the case: e.g., two different choices of the coefficient fields $(\gamma,\mathcal{A})$ lead in general to two inequivalent TRD--1 equations.  (See Section 3.4 in Ref. \cite{A42}.)  The cause of this variety is the variety of different choices for the coefficient fields $(\gamma,\mathcal{A})$, not the existence of two different representations (QRD and TRD) nor the variety of the possible connections. However, any two DFW equations on a given curved spacetime are equivalent for simple topologies of the spacetime \cite{Isham1978}. This is due to the fact that DFW restricts the choice of the coefficient fields $(\gamma,\mathcal{A})$ by expressing them from a tetrad field.
\footnote{\
In fact, the equivalence classes of DFW equations are in one-to-one correspondence with the homotopy classes of the tetrad fields \cite{Isham1978}.  A unique equivalence class exists if the spacetime is simply connected.
}\\

As a consequence of \hyperref[Theorem1]{Theorem 1} and \hyperref[Theorem2]{Theorem 2}, the DFW equation with any choice of the tetrad field is equivalent to some normal TRD--1 equation (\hyperref[Theorem3]{Theorem 3}).  [Here, ``normal" refers to the usual Dirac equation (\ref{Dirac-general}), in contrast with the modified one (\ref{Dirac-general-modified}).] That is, any DFW equation is equivalent to a particular case of the normal TRD--1 equation, that particular case being obtained by choosing the coefficient fields inside a special class of all possible coefficient fields for the normal TRD--1 equation. In short: the linear normal TRD--1 equation generalizes the DFW equation. It follows that TRD--1 can describe spin-half particles in a curved spacetime as well as DFW can.  \\

\vspace{3mm}
{\bf Acknowledgement.} We are grateful to Professor Gennadi Sardanashvily for his detailed questions which allowed us to significantly improve the clarity in the presentation of several crucial points.

\appendix
\section{Appendix: Proof that $\mathrm{V} \times{\sf C}^4$ and $\mathrm{T}_{\sf C}\mathrm{V}$ are spinor bundles}\label{Construc-gamma}

In this Appendix we present an explicit construction of a Dirac gamma field $\gamma $
and a hermitizing metric $(\ ,\ )$, for any parallelizable complex four-dimensional
vector bundle (e.g., $\mathrm{V} \times{\sf C}^4$ or $\mathrm{T}_{\sf C}\mathrm{V}$) over a parallelizable spacetime $(\mathrm{V}, <\ ,\ > )$. We begin with a straightforward lemma:\\

{\bf Lemma A.} (i) {\it The trivial bundle $\mathrm{V} \times{\sf C}^4$ is parallelizable.} (ii) {\it Suppose that there exists a global frame field $(u_\alpha )$ on the tangent bundle $\mathrm{TV}$. Then the complex tangent bundle $\mathrm{T}_{\sf C}\mathrm{V}$ is parallelizable.}\\

{\it Proof.} Recall that, by definition, a vector bundle $\mathrm{E}$ with base $\mathrm{V} $ is parallelizable if and only if there exists at least one frame field $X \mapsto (e_a(X))$ that is defined globally, i.e., for every $X \in \mathrm{V}$. (i) It is clear that the canonical basis $(E_a)$ of ${\sf C}^4$ defines a (constant) global frame field $X \mapsto (e_a(X))\equiv (E_a)$ on $\mathrm{V} \times{\sf C}^4$. (ii) Assume that $(u_\alpha )$ is a global frame field on the tangent bundle $\mathrm{TV}$. Let us define the following frame field on the complex tangent bundle $\mathrm{T}_{\sf C}\mathrm{V}$:
\be
e_a=\delta ^\alpha _a u_\alpha .
\ee 
Clearly, this is a global frame field on $\mathrm{T}_{\sf C}\mathrm{V}$. $\hspace {55mm} \square$\\

\vspace{4mm}
{\bf Theorem A.} {\it Assume that there exists an orthonormal tetrad field $(u_\alpha)$ that is defined globally in the spacetime $(\mathrm{V},<\,,\,>)$. Let $\mathrm{E}$ be a four-dimensional complex vector bundle over $\mathrm{V}$.  Suppose that $\mathrm{E}$ is parallelizable. Then $\mathrm{E}$ is a spinor bundle.}\\

{\it Proof.} Since $\mathrm{E}$ is parallelizable, there exists a global frame field $(e_a)$ on $\mathrm{E}$. Let $(\theta^a)$ be the dual frame field of $(e_a)$, which is a global frame field on the dual vector bundle $\mathrm{E}^\circ$ of E. Let $\gamma^{\alpha a } _{ b}=\left(C^\alpha \right)^a  _{\ \ b}$ be the components of constant matrices $C^\alpha$ acting on ${\sf C}^4$, satisfying the anticommutation relation:
\be \label{Clifford flat}
C^\alpha C^\beta + C^\beta C^\alpha   = 2\eta^{\alpha\beta} \,{\bf 1}_4, 
\ee
where $\eta^{\alpha\beta}$ is Minkowski metric, and ${\bf 1}_4$ is the identity matrix on ${\sf C}^4$.  The  constant matrices $C^\alpha$ satisfying Eq. (\ref{Clifford flat}) can be chosen to be just the standard constant Dirac gamma matrices for the Minkowski metric $\eta^{\alpha\beta}$, which we know satisfy Eq. (\ref{Clifford flat}).\\  

Let $p \in \mathrm{TV}$, thus $p \in \mathrm{TV}_X$ for some $X \in \mathrm{V}$, and let us define
\footnote{\
Recall that, as a set, the vector bundle $Hom(\mathrm{E},\mathrm{F})$ (where E and F are two vector bundles having a common base manifold V) is defined to be the union of the vector spaces 
$Hom(\mathrm{E}_X,\mathrm{F}_X)$ for $X \in \mathrm{V}$. See e.g. Ref. \cite{DieudonnéTome3}, Section 16.16.
}
\be\label{Def not p}
{\not}p \equiv  p_\alpha \gamma^{\alpha a } _{ b} \, e_a(X) \otimes \theta^b(X) \in Hom(\mathrm{E}_X,\mathrm{E}_X) \subset Hom(\mathrm{E},\mathrm{E}),
\ee
where $p_\alpha \equiv \eta _{\alpha \beta }\,p^\beta $ are the components, in the dual frame field of $(u_\alpha )$, of the covector $p^\flat$ which is associated with the vector $p$ by using the metric. Clearly, the map
\be
\mathrm{TV}\rightarrow Hom(\mathrm{E},\mathrm{E}), \qquad p \mapsto {\not}p
\ee
is indeed a smooth vector bundle map over Id$_\mathrm{V}$ as required by the definition of a spinor bundle \hyperref[DefSpinorBundle]{in Section 2.1}. \\

\vspace{3mm}
It remains to prove the coordinate-free anticommutation relation (\ref{Clifford slash}). To prove that relation, we may consider the components ${\not}p^a_{\ \,b}$ and ${\not}k^a_{\ \,b}$ of ${\not}p, {\not}k \in Hom(\mathrm{E},\mathrm{E})$ in the frame fields $(e_a)$ and $(\theta ^b)$, which are apparent in Eq. (\ref{Def not p}) above. Thus we have:
\bea
\left({\not}\,p{\not}k+{\not}k\,{\not}p \right)^a_{\ \,c} & = & {\not}p^a_{\ \,b}\,{\not}k^b_{\ \,c}+{\not}k^a_{\ \,b}\,{\not}p^b_{\ \,c}\\
& = & p_\alpha \gamma^{\alpha a } _{ b}\,k_\beta  \gamma^{\beta  b } _{ c}+k_\alpha \gamma^{\alpha a } _{ b}\,p_\beta  \gamma^{\beta  b } _{ c}\\
& = & p_\alpha \,k_\beta\left(\gamma^{\alpha a } _{ b} \gamma^{\beta  b } _{ c} +\gamma^{\beta  a } _{ b} \gamma^{\alpha   b } _{ c}\right)\\
& = & p_\alpha \,k_\beta (2\eta ^{\alpha \beta }\,\delta ^a_c)\\
& = & 2 <p,k>\,\delta ^a_c,
\eea
Q.E.D.\\

\vspace{5mm}
{\bf Corollary A.} {\it Assume that there exists an orthonormal tetrad field $(u_\alpha)$ that is defined globally in the spacetime $(\mathrm{V},<\,,\,>)$. Then both the trivial bundle $\mathrm{V} \times{\sf C}^4$ and the complex tangent bundle $\mathrm{T}_{\sf C}\mathrm{V}$ are spinor bundles.}\\

{\it Proof.}  This is an immediate consequence of Lemma A and Theorem A.\\

\vspace{3mm} As summarized by Eq. (\ref{gamma bundle}), a smooth vector bundle map $\mathrm{TV}\rightarrow Hom(\mathrm{E},\mathrm{E})$ can be regarded as a section of $\mathrm{TV}	 \otimes  \mathrm{E} \otimes \mathrm{E}^\circ$, or ``$\gamma$ field". Thus Theorem A guarantees the existence of at least one $\gamma $ field, which is such that, in any local chart $\chi: \mathrm{V}\supset \mathrm{U} \rightarrow {\sf R}^4$, the associated Dirac matrices $\gamma^\mu$, as in Eq. (\ref{gamma^mu from gamma intrinsic}), satisfy the anticommutation relation (\ref{Clifford}) in the curved spacetime $(\mathrm{U},<\,,\,>)$. [In the frame fields: $(u_\alpha ),\,(e_a),\,(\theta ^b)$ considered in the proof of Theorem A, the components of $\gamma $ are $\gamma^{\alpha a } _{ b}=\left(C^\alpha \right)^a  _{\ \ b}$.] Note that a similar construction holds for the global hermitizing metric $(\,,\,)$, provided that the matrix $C^0$ is hermitizing for the matrices $C^\alpha $. 


\end{document}